\documentclass[conference]{IEEEtran}
\IEEEoverridecommandlockouts
\usepackage{cite}
\usepackage{amsmath,amssymb,amsfonts}
\usepackage{algorithmic}
\usepackage{graphicx}
\usepackage{textcomp}
\usepackage{xcolor}
\usepackage{booktabs}
\usepackage{multirow}
\usepackage{tabularx}
\usepackage{hyperref}
\usepackage{url}
\usepackage{subfig}
\usepackage{tikz}
\usepackage{enumitem}
\def\BibTeX{{\rm B\kern-.05em{\sc i\kern-.025em b}\kern-.08em
    T\kern-.1667em\lower.7ex\hbox{E}\kern-.125emX}}

\newcommand{\circnum}[1]{\textcircled{\scriptsize #1}}

\newcommand\copyrighttext{%
  \footnotesize This work has been submitted to the IEEE for possible publication. Copyright may be transferred without notice, after which this version may no longer be accessible.}
\newcommand\copyrightnotice{%
\begin{tikzpicture}[remember picture,overlay]
\node[anchor=south,yshift=11pt] at (current page.south) {\fbox{\parbox{\dimexpr\textwidth-\fboxsep-\fboxrule\relax}{\copyrighttext}}};
\end{tikzpicture}%
}
    
\begin{document}

\title{Security Risks in Machining Process Monitoring: Sequence-to-Sequence Learning for Reconstruction of CNC Axis Positions}

\author{\IEEEauthorblockN{Lukas Krupp, Rickmar Stahlschmidt and Norbert Wehn}
\IEEEauthorblockA{RPTU University Kaiserslautern-Landau, Kaiserslautern, Germany}
}

\maketitle

\maketitle
\IEEEpubid{\begin{minipage}{\textwidth}
  \copyrightnotice
\end{minipage}} 

\begin{abstract}
Accelerometer-based process monitoring is widely deployed in modern machining systems. When mounted on moving machine components, such sensors implicitly capture kinematic information related to machine motion and tool trajectories. If this information can be reconstructed, condition monitoring data constitutes a severe security threat, particularly for retrofitted or weakly protected sensor systems. Classical signal processing approaches are infeasible for position reconstruction from broadband accelerometer signals due to sensor- and process-specific non-idealities, like noise or sensor placement effects.
In this work, we demonstrate that sequence-to-sequence machine learning models can overcome these non-idealities and enable reconstruction of CNC axis and tool positions. Our approach employs LSTM-based sequence-to-sequence models and is evaluated on an industrial milling dataset.
We show that learning-based models reduce the reconstruction error by up to $\mathbf{98\%}$ for low complexity motion profiles and by up to $\mathbf{85\%}$ for complex machining sequences compared to double integration. Furthermore, key geometric characteristics of tool trajectories and workpiece-related motion features are preserved.
To the best of our knowledge, this is the first study demonstrating learning-based CNC position reconstruction from industrial condition monitoring accelerometer data.

\end{abstract}

\begin{IEEEkeywords}
CNC machining, process monitoring, MEMS accelerometers, sequence-to-sequence learning, LSTMs, position reconstruction
\end{IEEEkeywords}

\section{Introduction}
Sensor-based machining process monitoring has become a necessity in modern manufacturing due to continuously increasing quality and efficiency demands \cite{b1}. In many industrial sectors, even minor deviations in machining quality can result in significant economic losses, safety risks, or performance degradation of the final products. Examples include:
\begin{itemize}
    \item electric vehicles, where the absence of engine noise makes mechanical irregularities perceptible to the user,
    \item wind turbines, where bearing surface quality directly affects efficiency and lifetime, 
    \item mold making, where the surface quality of molds defines the quality of mass-produced plastic components.
\end{itemize}

Across these domains, vibration and structure-borne sound are among the most informative signals. As a result, accelerometer-based sensing has become a de facto standard for tool condition monitoring, chatter detection, surface quality estimation, and process supervision \cite{b2, b3, b4}. Accelerometers are inexpensive, compact, and easily retrofittable. Broadband micro-electro-mechanical systems (MEMS) accelerometers with low noise density, such as the Analog Devices \textit{ADXL100x} family, are increasingly used to capture the wide frequency range induced by high-speed machining \cite{b5}.

Although these accelerometers are not intended for determining machine or tool position, their signals implicitly encode more information than tool condition alone. When mounted on moving machine components such as the spindle, the measured acceleration reflects not only process-induced vibrations but also axis motion and tool trajectories. In principle, position information can be obtained by double integration of acceleration signals \cite{b6}. However, due to integration drift, classical signal processing approaches are unsuitable for accurate position reconstruction from industrial accelerometer data.

This assumption is challenged by recent advances in sequence-to-sequence machine learning (ML) \cite{b7}. In domains such as human motion tracking, data-driven approaches, particularly recurrent neural networks such as Long Short-Term Memory (LSTM) models, have demonstrated the ability to reconstruct velocity and position information from noisy, low-bandwidth acceleration signals. These models implicitly learn to compensate for sensor imperfections, noise accumulation, and drift effects that ultimately cause integration drift \cite{b8}.

Transferring these capabilities to machining process monitoring data raises a critical and largely unexplored concern. In manufacturing domains such as mold making, automotive, and aerospace production, machining strategies, tool paths, and workpiece geometries constitute highly sensitive intellectual property (IP) and a key competitive advantage. If machine learning models are able to reconstruct position data or workpiece-related motion features from accelerometer signals, access to such data, e.g., via compromised wireless sensor links or edge devices \cite{b9}, could enable reverse engineering of protected manufacturing knowledge.

In this paper, we demonstrate that such sensor data can be exploited to retrieve critical manufacturing IP and and thereby pose a significant security risk. Our key contributions are:
\begin{itemize}
    \item A data-driven approach for reconstructing CNC motion trajectories from spindle-mounted broadband accelerometer data acquired for milling process monitoring.
    \item A quantitative comparison between classical double integration and LSTM-based sequence-to-sequence models under realistic industrial noise and drift conditions.
    \item An evaluation across milling scenarios with increasing motion and process complexity, and learning formulations (many-to-one, many-to-many, autoregressive).
\end{itemize}

\section{Background and Related Work} \label{sec:related_work}
\begin{figure}[htbp]
\centerline{\includegraphics[width=0.5\textwidth]{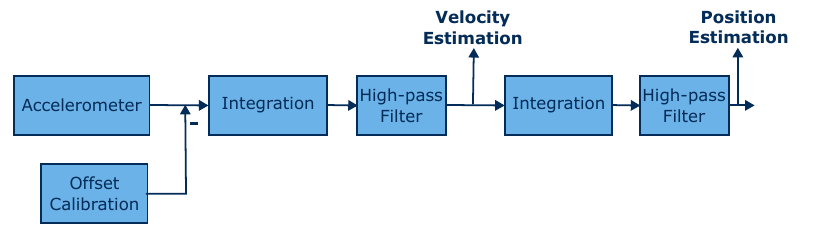}}
\caption{State-of-the-art digital signal processing approach to improve double integration for position estimation \cite{b6}.}
\label{fig:integration_scheme}
\end{figure}

The classical approach for reconstructing position from acceleration measurements is based on numerical double integration \cite{b6}. This pipeline typically consists of sensor offset calibration, anti-aliasing filtering, numerical integration, and high-pass filtering to suppress integration drift as shown in Figure~\ref{fig:integration_scheme}. Offset calibration is essential, as residual $0g$ bias is quadratically amplified during double integration, leading to severe position drift. Anti-aliasing low-pass filters are often already implemented within industrial sensor systems and are therefore assumed as part of the signal acquisition chain.

Despite these preprocessing steps, numerical integration remains highly sensitive to noise and low-frequency disturbances. High-pass filtering after each integration stage is commonly applied to mitigate drift, but introduces a critical trade-off between drift suppression and preservation of physically meaningful motion components. Consequently, integration-based approaches require careful filter tuning and remain fundamentally limited for long observation windows and noisy broadband acceleration signals.

To overcome these limitations, recent research has increasingly explored sequence-to-sequence ML models that directly learn the mapping from acceleration sequences to kinematic quantities. Most existing work focuses on human motion tracking \cite{b8, b10}, inertial navigation \cite{b11, b12, b13}, or orientation estimation \cite{b7, b14}, employing architectures such as convolutional neural networks, multilayer perceptrons, LSTMs, gated recurrent units, temporal convolutional networks, transformers, and state-space models. Across these domains, recurrent architectures and LSTMs in particular demonstrate strong or best-in-class performance for sequence regression tasks.

The literature overview reveals a clear research gap: previous work does not address position reconstruction from condition monitoring accelerometer data. In contrast to inertial measurement units (IMUs), such sensors capture a superposition of kinematic motion, process-induced vibrations, and structural resonances. To the best of our knowledge, learning-based position reconstruction from this type of accelerometer data has not been previously investigated.

\section{Methodology}
\begin{figure*}[htbp]
  \centering
  \includegraphics[width=\linewidth]{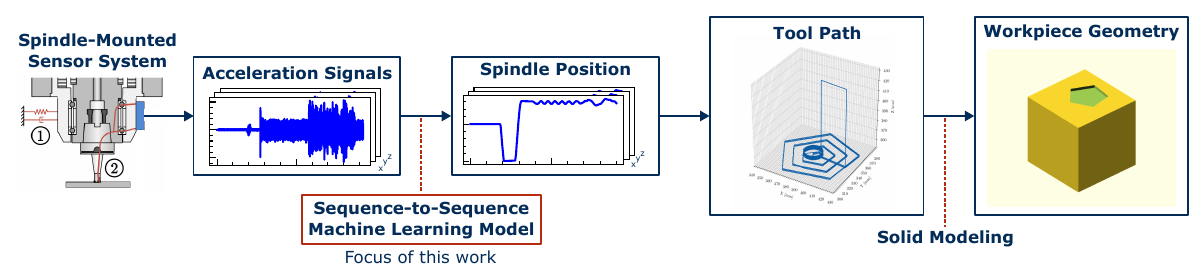}
  \caption{Conceptual information extraction pipeline illustrating how process monitoring data can be transformed into higher-level representations, potentially leading to unintended leakage of sensitive process and design information.}
  \label{fig:methodology_1}
\end{figure*}

Figure~\ref{fig:methodology_1} illustrates the information extraction pipeline considered in this work, highlighting how seemingly low-level process monitoring data can be transformed into abstract and sensitive representations. 
Starting from a spindle-mounted acceleration sensor system, raw broadband multi-axis acceleration signals are acquired during machining. These signals inherently encode a superposition of process-induced vibrations and kinematic motion components originating from the machine tool drives \circnum{1} and process-induced structure-borne sound \circnum{2}. By applying sequence-to-sequence ML models, the acceleration time series is mapped to spindle position trajectories, which represent a higher semantic level of information. From these trajectories, tool paths can be inferred, ultimately enabling the recovery of workpiece geometry through state-of-the-art solid modeling techniques \cite{b15}.

\begin{figure}[htbp!]
\centerline{\includegraphics[width=0.5\textwidth]{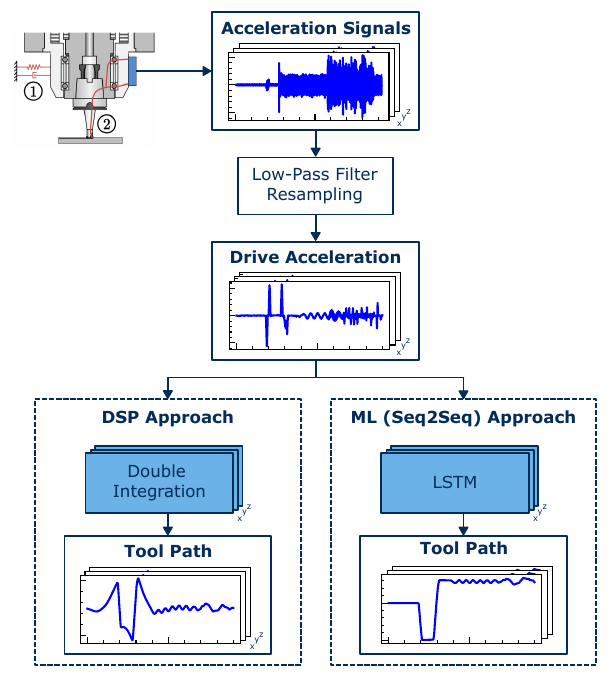}}
\caption{Overview of the two evaluated position reconstruction approaches.}
\label{fig:methodology_2}
\end{figure}

Figure~\ref{fig:methodology_2} provides an overview of the two position reconstruction approaches evaluated in this work: a classical digital signal processing (DSP) pipeline based on numerical integration, and an ML pipeline using sequence-to-sequence (Seq2Seq) models. Both approaches operate on the same acceleration signals and aim to recover the tool path corresponding to the kinematic motion of the machine tool axes.

The raw spindle-mounted accelerometer signal is denoted as $\mathbf{a}_{\mathrm{raw}}(t)\in\mathbb{R}^3$ and represents a superposition of multiple physical effects,
\begin{equation}
\mathbf{a}_{\mathrm{raw}}(t)
=
\mathbf{a}_{\mathrm{kin}}(t)
+
\mathbf{a}_{\mathrm{proc}}(t)
+
\mathbf{b}
+
\boldsymbol{\eta}(t),
\end{equation}
where $\mathbf{a}_{\mathrm{kin}}(t)$ captures the kinematic motion induced by the machine tool drives, $\mathbf{a}_{\mathrm{proc}}(t)$ contains process-induced vibrations and structure-borne sound, $\mathbf{b}$ is a quasi-static sensor bias, and $\boldsymbol{\eta}(t)$ denotes measurement noise. The signal is sampled at frequency $f_{\mathrm{raw}}$, yielding the discrete-time sequence $\mathbf{a}_{\mathrm{raw}}[n]$. The separation of kinematic and process-induced components is consistent with established models of machining vibration signals used in tool condition monitoring literature \cite{b16}.

A key assumption underlying both pipelines is the availability of domain-specific prior knowledge about the machine tool dynamics. Due to the inertia of the machine tool and bounded axis velocities, the kinematic acceleration $\mathbf{a}_{\mathrm{kin}}(t)$ is inherently low-frequency, whereas process-induced vibrations and structure-borne sound predominantly occupy higher frequency bands. Since this work explicitly targets the reconstruction of kinematic motion, this separation is exploited through frequency-domain preprocessing.

Accordingly, the raw acceleration signal is low-pass filtered using a filter $h_{\mathrm{LP}}$ with cut-off frequency $f_c$ and subsequently resampled,
\begin{equation}
\tilde{\mathbf{a}}[n] = (h_{\mathrm{LP}} * \mathbf{a}_{\mathrm{raw}})[n],
\end{equation}
\begin{equation}
\mathbf{a}[k] = \tilde{\mathbf{a}}[rk], \quad r=\frac{f_{\mathrm{raw}}}{f},
\end{equation}
where $f_{\mathrm{raw}}$ and $f$ denote the original and reduced sampling rates, respectively. The resulting discrete-time signal $\mathbf{a}[k]$ is referred to as drive acceleration and constitutes the common input to both reconstruction pipelines. The objective of both approaches is to estimate the spindle position trajectory $\mathbf{p}[k]\in\mathbb{R}^3$, $k=0,\dots,N-1$, from the preprocessed acceleration sequence $\mathbf{a}[0{:}N-1]$.

In the DSP approach \cite{b6}, position reconstruction relies on the physical relationship between acceleration, velocity, and position. After preprocessing, the drive acceleration is numerically integrated to estimate velocity,
\begin{equation}
\hat{\mathbf{v}}[k]
=
\hat{\mathbf{v}}[k-1]
+
\frac{T}{2}\left(\mathbf{a}[k]+\mathbf{a}[k-1]\right),
\end{equation}
where $T=1/f$ is the sampling interval. A second integration yields the position estimate,
\begin{equation}
\hat{\mathbf{p}}_{\mathrm{int}}[k]
=
\hat{\mathbf{p}}_{\mathrm{int}}[k-1]
+
\frac{T}{2}\left(\hat{\mathbf{v}}[k]+\hat{\mathbf{v}}[k-1]\right).
\end{equation}
However, sensor bias and noise components are amplified through integration, leading to drift. To mitigate this effect, high-pass filtering is applied after each integration stage,
\begin{equation}
\hat{\mathbf{v}}_{\mathrm{HP}}[k]=\mathrm{HP}_v(\hat{\mathbf{v}}[k]), \quad
\hat{\mathbf{p}}_{\mathrm{rel}}[k]=\mathrm{HP}_p(\hat{\mathbf{p}}_{\mathrm{int}}[k]).
\end{equation}
The final absolute position estimate is obtained by adding the known initial tool position $\mathbf{p}_0$,
\begin{equation}
\hat{\mathbf{p}}_{\mathrm{DSP}}[k]=\mathbf{p}_0+\hat{\mathbf{p}}_{\mathrm{rel}}[k].
\end{equation}

In contrast to explicit numerical integration, the ML-based approach formulates position reconstruction as a direct regression problem on temporal sequences. Formally, the task is to learn a parametric mapping
\begin{equation}
\hat{\mathbf{p}}_{\mathrm{ML}}[0{:}N-1]
=
f_{\boldsymbol{\theta}}\!\left(\mathbf{a}[0{:}N-1]\right),
\end{equation}
where $f_{\boldsymbol{\theta}}$ is implemented by a LSTM network with parameters $\boldsymbol{\theta}$. LSTMs are chosen due to their state-of-the-art performance in time-series modeling \cite{b17} and their successful application to position reconstruction and inertial navigation tasks, as discussed in Section~\ref{sec:related_work}. Their ability to capture long-range temporal dependencies and internal state evolution makes them well suited for modeling the cumulative effects inherent in acceleration-to-position transformations. LSTMs serve as a basis for constructing different sequence-to-sequence learning formulations, which are discussed in Section~\ref{sec:learning_formulations}.

\section{Experimental Setup and Implementation}
\subsection{Dataset and Sensor System}\label{subsec:dataset_sensor_system}
The experimental evaluation in this work is based on an existing milling dataset that was originally collected for tool condition monitoring and remaining tool life prediction \cite{b16}. The dataset was acquired on a five-axis CNC milling center during pocket- and face-milling operations. For each experiment, geometric pocket parameters and cutting parameters were sampled, followed by the execution of consecutive pocket milling operations interleaved with intermittent face milling. Tool wear and surface roughness measurements were performed at predefined intervals, and each experiment was terminated once a specified flank wear threshold was exceeded.

Vibration and structure-borne sound signals were recorded using a spindle-mounted accelerometer system consisting of three single-axis \textit{ADXL1002} broadband MEMS accelerometers \cite{b5}. The sensors were rigidly attached to the spindle housing and aligned with the machine’s Cartesian axes, enabling the measurement of translational accelerations along the $x$-, $y$-, and $z$-directions. This placement allows continuous monitoring without interfering with the machining process.

The accelerometers were operated at a sampling rate of 256~kHz. In parallel, ground-truth axis position signals were obtained from the machine controller at a frequency of 110~Hz. Synchronization between the sensor data streams was ensured during preprocessing, enabling accurate temporal alignment between acceleration inputs and position targets. The resulting dataset comprises multiple experimental series with varying pocket geometries, tool paths, and cutting parameters, thereby covering a wide range of kinematic motion profiles.

\subsection{Data Preparation}\label{subsec:data_preparation}
\begin{figure}[htbp]
\centerline{\includegraphics[width=0.5\textwidth]{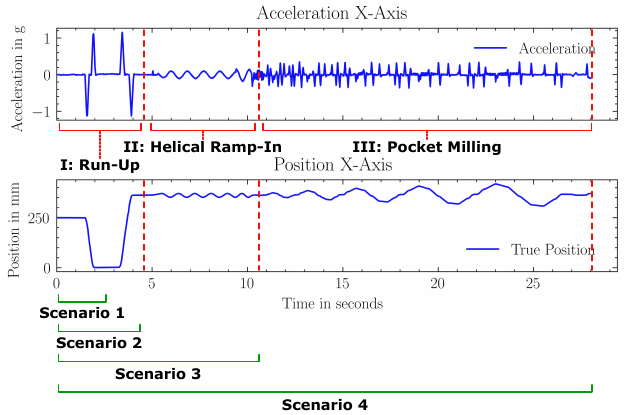}}
\caption{Exemplary milling sequence illustrating the relationship between machining phases and the four evaluated prediction scenarios. Only the $x$-axis components of the acceleration and ground-truth position signals are shown.}
\label{fig:scenarios}
\end{figure}

In this work, only the synchronized acceleration signals and the corresponding ground-truth axis position trajectories are used as shown in Fig.~\ref{fig:scenarios}. 

From the full recordings, contiguous temporal sequences are extracted. The dataset is split at the level of milling experiments to avoid information leakage between training and testing data. The training dataset comprises six independent series of milling experiments, each contributing six machining sequences, resulting in a total of 36 training sequences. The testing dataset consists of two separate experiment series, again with six sequences per series, yielding 12 test sequences. No sequences from the same experimental series appear in both subsets. Each sequence reflects the underlying machining process and can be divided into three characteristic phases:
\begin{enumerate}[label=\Roman*)]
    \item \textbf{Run-up}: The tool is moved from a fixed initial parking position to the starting point of the machining operation. This phase is dominated by deterministic acceleration and deceleration patterns and exhibits low variability across different sequences.
    \item \textbf{Helical ramp-in}: The tool enters the material along a helical trajectory. In this phase, circular motions introduce more complex, quasi-periodic acceleration components.
    \item \textbf{Pocket milling}: The actual material removal is performed. This phase is characterized by highly variable and process-dependent acceleration signatures caused by tool–workpiece interaction, cutting dynamics, and changing tool paths.
\end{enumerate}

To systematically analyze the influence of motion complexity on position reconstruction performance, four scenarios with increasing difficulty are defined by selecting different temporal subsets of each machining cycle:
\begin{itemize}
    \item \textbf{Scenario~1}: The first half of the run-up phase only. This scenario represents the simplest case, with nearly identical motion profiles across all sequences.
    \item \textbf{Scenario~2}: The complete run-up phase. Compared to Scenario~1, the duration is increased and the end positions vary, introducing moderate variability.
    \item \textbf{Scenario~3}: Run-up followed by the helical ramp-in phase. This scenario introduces circular motion components and significantly increases complexity.
    \item \textbf{Scenario~4}: The complete leading part of the machining cycle, including run-up, ramp-in, and pocket milling. This represents the most challenging setting, as it combines long observation windows with highly diverse and process-dependent acceleration patterns.
\end{itemize}

These scenarios enable the evaluation of both classical signal processing and sequence-to-sequence learning approaches under progressively more demanding conditions.

\subsection{Learning Formulations and Implementation}\label{sec:learning_formulations}
Three sequence-to-sequence learning formulations are considered. In the many-to-one formulation, a fixed-length window of acceleration samples is mapped to it's last position value. In the many-to-many formulation, a complete acceleration sequence is mapped to a position sequence of equal length. Since the milling sequences exhibit varying durations, zero-padding is applied to enable batch processing. In the autoregressive formulation, the model predicts the position at the next timestep based on the current acceleration input and the previous position. During training, ground-truth position values are provided and during inference, the predicted position values are recursively fed back into the model.

The learning approaches are based on an LSTM model consisting of two stacked LSTM layers followed by a fully connected output layer, conceptually mirroring the two-stage integration. Separate models are trained for each spatial axis. All models are implemented in \textit{PyTorch} and trained using the Adam optimizer with a mean squared error loss. Hyperparameters are optimized using Bayesian optimization.

The DSP-based position reconstruction pipeline is implemented using the \textit{SciPy} library. After preprocessing, the acceleration signals are integrated using cumulative trapezoidal numerical integration to obtain velocity and position estimates. After each integration stage zero-phase digital high-pass filtering is applied. The cutoff frequencies and filter orders of the high-pass filters impact the trade-off between drift suppression and the preservation of physically meaningful motion components. The filter parameters are optimized using a black-box optimization framework \cite{b18}. 

The trained LSTM models, the hyperparameters, and the compensated double-integration approach including the filter parameters are publicly available\footnote{Repository: \url{https://github.com/tukl-msd/seq2seq-position-reconstruction}}.

\section{Results}
\subsection{Quantitative Reconstruction Accuracy}
Table~\ref{tab:3d_error_all_scenarios_compact} summarizes the three-dimensional position reconstruction errors $\lVert\mathbf{p}\rVert$ for all evaluated approaches and scenarios, reported as mean $\pm$ standard deviation in millimeters. The error is computed as the Euclidean norm of the instantaneous position error vector, thereby aggregating deviations along the $x$-, $y$-, and $z$-axes. Both the mean absolute error (MAE) and the root mean squared error (RMSE) are reported to characterize average reconstruction accuracy as well as sensitivity to larger deviations. All values are averaged over the complete test set of each scenario.

\begin{table*}[t]
\centering
\caption{3D position reconstruction error $\|\mathbf{p}\|$ for all scenarios
(mean $\pm$ std, in mm).}
\label{tab:3d_error_all_scenarios_compact}
\footnotesize
\setlength{\tabcolsep}{6pt}
\renewcommand{\arraystretch}{1.2}

\begin{tabular}{l l c c c c}
\toprule
\textbf{Variation} & \textbf{Metric}
& \textbf{Scen.\ 1}
& \textbf{Scen.\ 2}
& \textbf{Scen.\ 3}
& \textbf{Scen.\ 4} \\
\midrule

\multirow{2}{*}{Double Integration}
& MAE
& 201.89$\pm$22.16
& 244.13$\pm$52.49
& 279.19$\pm$40.45
& 344.18$\pm$71.83 \\
& RMSE
& 273.99$\pm$18.02
& 271.31$\pm$55.80
& 300.75$\pm$44.08
& 358.29$\pm$76.27 \\
\midrule

\multirow{2}{*}{Many-to-One}
& MAE
& 11.52$\pm$2.03
& 43.33$\pm$12.79
& 59.19$\pm$17.45
& 84.82$\pm$14.35 \\
& RMSE
& 17.02$\pm$4.62
& 56.40$\pm$19.53
& 66.39$\pm$18.93
& 88.14$\pm$14.08 \\
\midrule

\multirow{2}{*}{Many-to-Many}
& MAE
& 5.26$\pm$0.62
& 22.46$\pm$5.77
& 52.59$\pm$10.34
& 63.41$\pm$11.29 \\
& RMSE
& 10.62$\pm$0.73
& 35.09$\pm$8.69
& 64.75$\pm$12.76
& 70.78$\pm$12.43 \\
\midrule

\multirow{2}{*}{\textbf{Autoregressive}}
& MAE
& \textbf{2.68$\pm$0.30}
& \textbf{15.96$\pm$5.96}
& \textbf{25.12$\pm$8.95}
& \textbf{49.45$\pm$11.13} \\
& RMSE
& \textbf{4.77$\pm$0.62}
& \textbf{24.18$\pm$9.32}
& \textbf{29.04$\pm$10.49}
& \textbf{55.42$\pm$12.05} \\
\bottomrule
\end{tabular}
\end{table*}

The classical double-integration baseline exhibits large reconstruction errors across all scenarios, with RMSE values increasing from $273.99\,\mathrm{mm}$ in Scenario~1 to $358.29\,\mathrm{mm}$ in Scenario~4. This behavior reflects the well-known error accumulation and drift effects inherent to numerical integration of noisy acceleration signals, which become increasingly dominant for longer observation intervals.

In contrast, all sequence-to-sequence approaches achieve a substantial reduction in reconstruction error. For the simplest motion pattern in Scenario~1, both the many-to-one and many-to-many variants already reduce the RMSE by more than one order of magnitude compared to the classical baseline, achieving $17.02\,\mathrm{mm}$ and $10.62\,\mathrm{mm}$, respectively. The autoregressive formulation yields the best performance, with an RMSE of only $4.77\,\mathrm{mm}$, indicating that leveraging temporal feedback enables highly accurate recovery of the tool trajectory.

As the scenario complexity increases, reconstruction errors grow for all learning-based approaches, reflecting increased variability in the motion profiles and longer prediction horizons. Nevertheless, the relative ranking of the models remains consistent across all scenarios. The autoregressive approach consistently achieves the lowest errors, followed by the many-to-many variant, while the many-to-one formulation shows the weakest performance among the learned models. In the most challenging case (Scenario~4), the autoregressive model attains an RMSE of $55.42\,\mathrm{mm}$, compared to $70.78\,\mathrm{mm}$ for many-to-many and $88.14\,\mathrm{mm}$ for many-to-one. All learning-based approaches outperform classical double integration.

\subsection{Trajectory-Level Analysis and Interpretability}
\begin{figure*}[t]
    \centering

    \subfloat[Axis-wise position over time.]{
        \includegraphics[width=0.35\linewidth]{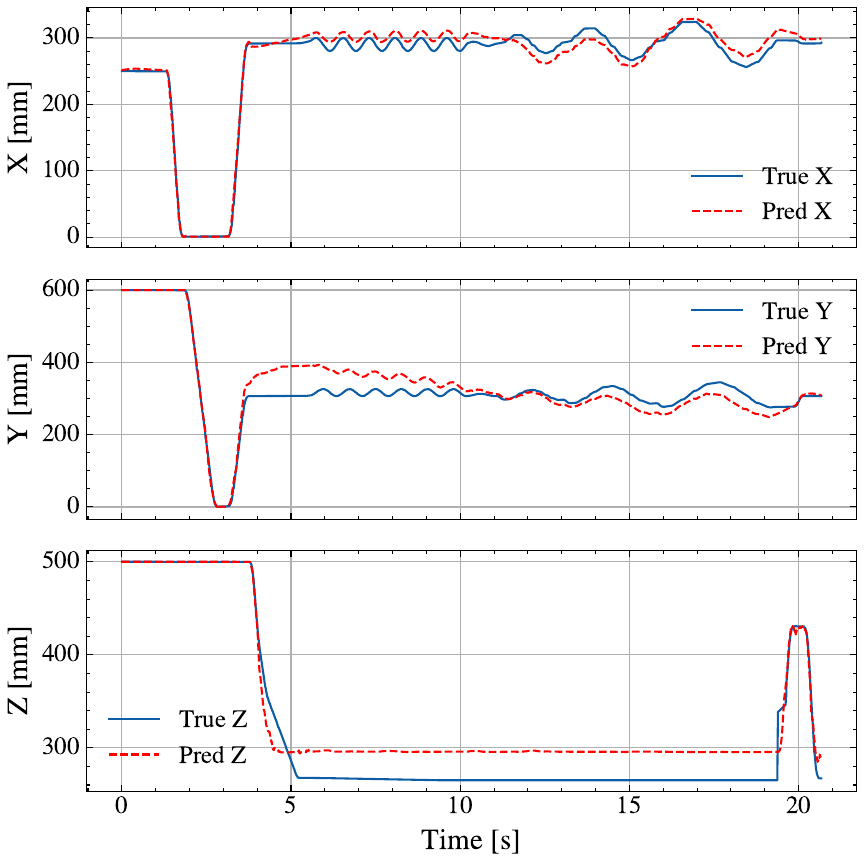}
        \label{fig:ts_xyz}
    }\hfill
    \subfloat[2D trajectory.]{
        \includegraphics[width=0.22\linewidth]{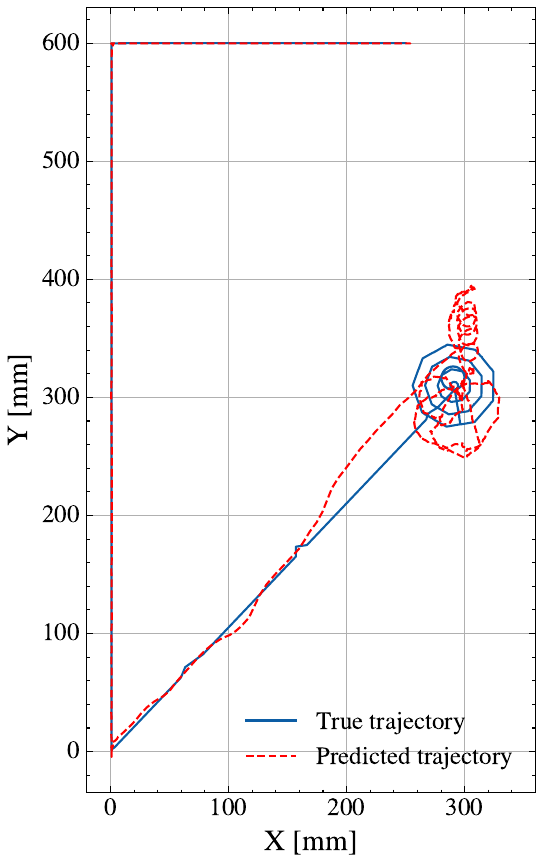}
        \label{fig:traj_xy}
    }\hfill
    \subfloat[3D trajectory.]{
        \includegraphics[width=0.35\linewidth]{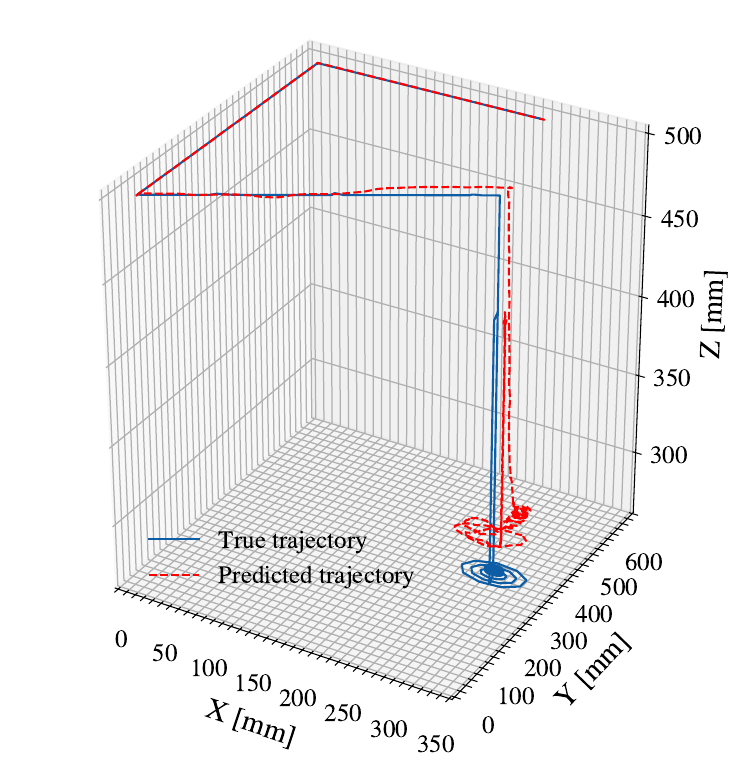}
        \label{fig:traj_3d}
    }

    \caption{Qualitative trajectory reconstruction for a representative test sequence.}\label{fig:qualitative_results}
\end{figure*}

Fig.~\ref{fig:qualitative_results} provides a qualitative comparison between the ground-truth position and the reconstructed trajectory for a representative test sequence. 
Fig.~\ref{fig:qualitative_results}(a) depicts the axis-wise position estimates over time. While the predicted signals closely follow the overall temporal evolution of the ground truth for all three axes, small deviations and phase shifts are observable. Nevertheless, the dominant motion patterns and transitions are captured reliably in the time domain.

The accumulated effect of these axis-wise errors becomes more apparent in the geometric trajectory representations shown in Fig.~\ref{fig:qualitative_results}(b) and (c). Although key structural features of the motion, such as straight segments and turning points, remain recognizable, the reconstructed trajectories exhibit distortions relative to the ground truth in both the 2D and the 3D view. In particular, local deviations and trajectory spreading are visible, indicating that small per-axis prediction errors compound when mapped into Cartesian space.

This behavior is attributed to the use of independent sequence-to-sequence models for each spatial axis. Since the $x$, $y$, and $z$ components are predicted separately, inter-axis correlations and geometric consistency are not explicitly enforced, leading to misalignment effects in the reconstructed trajectory despite reasonable axis-wise accuracy. A joint multi-output model that predicts all three spatial components simultaneously could leverage cross-axis dependencies and is therefore expected to further improve geometric consistency.

\section{Conclusion}
In this paper, we evaluated the potential to reconstruct CNC axis positions from broadband accelerometer data using sequence-to-sequence machine learning models.
Classical double integration fails under realistic industrial conditions, motivating the use of data-driven reconstruction approaches.
Accordingly, LSTM-based models were proposed and validated on an industrial milling dataset.
For low-complexity motion scenarios, the reconstruction error is reduced by up to $98\%$ relative to classical integration, while for complex machining sequences including ramp-in and pocket milling, the models outperform double integration by more than $85\%$.

The results indicate that accelerometer signals acquired for process monitoring implicitly encode high-level kinematic information that can be recovered. This has important implications for manufacturing security: vibration data from retrofitted or weakly protected sensor systems may enable reconstruction of workpiece features and machining strategies. Consequently, sensing data is a carrier of process-related IP, extending the traditional IT/OT security focus beyond machine controllers. Cryptographic protection and secure transmission of sensor data are essential elements of industrial sensing infrastructures. 

Future work will follow two directions. First, the findings motivate the development of secure, retrofit sensor systems for machining process monitoring.
Second, the limits of information extraction from vibration data will be further explored by developing more expressive reconstruction models. The dataset will be extended to additional machine types to assess cross-machine generalization. Furthermore, a comparison of learning-based models beyond LSTMs will be conducted, and joint multi-axis models will be investigated.

\end{document}